\documentclass[pra,floatfix,twocolumn,superscriptaddress]{revtex4-1}

\usepackage{graphicx,epstopdf}
\usepackage{amsmath}
\usepackage{hyperref}
\usepackage{amsfonts}
\usepackage{bbm}
\usepackage{bbold}
\usepackage{amssymb}
\usepackage{color}
\usepackage{latexsym}
\usepackage{times,txfonts}

\begin{document}

\title{Superadiabatic Controlled Evolutions and Universal Quantum Computation}

\author{Alan C. Santos}
\affiliation{Instituto de F\'{\i}sica, Universidade Federal Fluminense, Av. General Milton Tavares de Souza, Gragoat\'a, 24210-346,Niter\'oi, RJ, Brazil}
\email{alancs@if.uff.br}

\author{Marcelo S. Sarandy}
\affiliation{Instituto de F\'{\i}sica, Universidade Federal Fluminense, Av. General Milton Tavares de Souza, Gragoat\'a, 24210-346,Niter\'oi, RJ, Brazil}
\affiliation{Center for Quantum Information Science \& Technology and Ming Hsieh Department of Electrical Engineering, University of Southern California, Los Angeles, California 90089, USA}
\email{msarandy@if.uff.br}

\date{\today}

\begin{abstract}

Adiabatic state engineering is a powerful technique in quantum information and quantum control. 
However, its performance is limited by the adiabatic theorem of quantum mechanics. 
In this scenario, shortcuts to adiabaticity, such as provided by the superadiabatic theory, constitute 
a valuable tool to speed up the adiabatic quantum behavior. Here, we propose a superadiabatic 
route to implement universal quantum computation. Our method is based on the realization of piecewise 
controlled superadiabatic evolutions. Remarkably, they can be obtained by simple time-independent counter-diabatic 
Hamiltonians. In particular, we discuss the implementation of fast rotation gates and arbitrary n-qubit controlled 
gates, which can be used to design different sets of universal quantum gates. Concerning the energy cost of 
the superadiabatic implementation, we show that it is dictated by the quantum speed limit, providing an 
upper bound for the corresponding adiabatic counterparts.

\end{abstract}

\maketitle

\section{Introduction}

Quantum adiabatic processes are a powerful strategy to implement quantum state engineering, which aims at manipulating 
a quantum system to attain a target state at a designed time T. In the adiabatic scenario, the quantum system evolves under a 
sufficiently slowly-varying Hamiltonian, which prevents changes in the populations of the energy eigenlevels. In particular, if 
the system is prepared in an eigenstate $|n(0)\rangle$ of the Hamiltonian $H$ at a time $t=0$, it will evolve to the corresponding 
instantaneous eigenstate $|n(t)\rangle$ at later times. This transitionless evolution is ensured by the adiabatic theorem, 
which is one of the oldest and most explored tools in quantum mechanics~\cite{Born:28,Kato:50,Messiah:book} . 
The huge amount of applications of the adiabatic behavior has motivated renewed interest in the adiabatic theorem, 
which has implied in its rigorous formulation~\cite{Teufel:03, Ambainis:04, Tong:05, Jansen:07, Amin:09, Tong:10, Cao:13} as well as 
in new bounds for adiabaticity~\cite{Tong:07,Yu:14,Wang:15} . 
In quantum information 
processing, the adiabatic theorem is the basis for the methodology of adiabatic quantum computation (AQC)~\cite{Farhi:01} , 
which has been originally proposed as an approach for the solution of hard combinatorial search problems. More generally, 
AQC has been proved to be universal for quantum computing, being equivalent to the standard circuit model of 
quantum computation up to polynomial resource-overhead~\cite{Aharonov:04} . Moreover, it is a physically appealing approach, 
with a number of experimental implementations in distinct architectures, e.g., nuclear magnetic resonance~\cite{Steffen:03,Peng:08,Long:13} , 
ion traps~\cite{Richerme:13} , and superconducting flux quantum bits (qubits) through the D-Wave quantum 
annealer~\cite{Johnson:11,Boixo:13,Boixo:14} .  

Recently, the circuit model has been directly connected with AQC via hybrid approaches~\cite{Bacon:09,Hen:15} . Then, an adiabatic circuit can 
be designed based on the adiabatic realization of quantum gates, which allows for the translation of the quantum circuit to the AQC 
framework with no further resources required. In particular, it is possible to implement universal sets of quantum gates through 
controlled adiabatic evolutions (CAE)~\cite{Hen:15} . In turn, CAE are used to perform one-qubit and two-qubit gates, allowing for universality 
through the set of one-qubit rotations joint with an entangling two-qubit gate~\cite{Barenco:95,Nielsen:book} . However, since these processes 
are ruled by the adiabatic approximation, it turns out that each gate of the adiabatic circuit will be implemented within some fixed probability (for a finite evolution time). 
Moreover, the time for performing each individual gate will be bounded from below by the adiabatic time condition~\cite{Teufel:03, Ambainis:04, Tong:05, Jansen:07, Amin:09, Tong:10, Cao:13} . 
For a recent analysis on adiabatic control of quantum gates and its corresponding non-adiabatic errors, see Ref.~\cite{Martinis:14} .

In order to resolve the limitations of adiabaticity in the hybrid model, we propose here a general shortcut to 
CAE through simple time-independent counter-diabatic assistant Hamiltonians within the framework of the superadiabatic 
theory~\cite{Demirplak:03,Demirplak:05,Berry:09,Torrontegui:13} . The physical resources spent by this strategy will be governed by the  
quantum circuit complexity, but no adiabatic constraint will be required in the individual implementation of the quantum gates. 
Moreover, the gates will be deterministically implemented with probability one as long as decoherence effects can be avoided. 
In particular, we discuss the realization of rotation gates and arbitrary n-qubit controlled gates, which can be used to design 
different sets of universal quantum gates. This analog approach allows for fast implementation of individual gates, whose time 
consumption is only dictated by the quantum speed limit (QSL) (for closed systems, see Refs.~\cite{Mandelstam:45,Margolus:98,Giovannetti:03,Deffner:13}). 
Indeed, the time demanded for each gate will imply in an energy cost, which increases with the speed of the evolution. In this 
context, by analyzing the energy-time complementarity, we will show that the QSL provides an energy cost for superadiabatic 
evolutions that upper bounds the cost of adiabatic implementations.

\section{Adiabatic quantum circuits}

Let us begin by discussing the design of adiabatic quantum circuits as introduced by Hen~\cite{Hen:15} through the implementation 
of quantum gates via CAE. 

\subsection{Controlled adiabatic evolution}
In order to define quantum gates through CAE, we will introduce a discrete bipartite system ${\cal SA}$ associated with a Hilbert 
space ${\cal H}_{\cal S} \otimes {\cal H}_{\cal A}$. The system ${\cal SA}$ is composed by a target subsystem ${\cal S}$ and an 
auxiliary subsystem ${\cal A}$, whose individual Hilbert spaces ${\cal H}_{\cal S}$ and ${\cal H}_{\cal A}$ have dimensions $d_{\cal S}$ and $d_{\cal A}$, 
respectively. The dynamics of  ${\cal SA}$ will be governed by a Hamiltonian in the form~\cite{Hen:15} 
\begin{equation}
H\left( t\right) =f\left( t\right) \left[ \mathbbm{1}\otimes H^{\left( b\right) }%
\right] +g\left( t\right) \left[ \sum\nolimits_{k}P_{k}\otimes H_{k}^{\left(
f\right) }\right] ,  \label{sce.1.1}
\end{equation}%
where $f\left( 0\right) =g\left( 1\right) =1$, $g\left( 0\right) =f\left(
1\right) =0$, and $\left\{ P_{k}\right\}$ denotes a complete set of orthogonal
projectors over ${\cal S}$, so that they satisfy $P_{k}P_{m}=\delta _{km}P_{k}
$ and $\sum_{k}P_{k}=1$. Alternatively, we can write Eq.~(\ref{sce.1.1}) as 
\begin{equation}
H\left( t\right) =\sum\nolimits_{k}P_{k}\otimes H_{k}\left( t\right), 
\label{sce.1.2}
\end{equation}%
with $H_{k}\left( t\right) =g\left( t\right) H_{k}^{\left( f\right)}+f\left( t\right) H^{\left( b\right) }$ denoting a Hamiltonian 
that acts on ${\cal A}$. Suppose now that we prepare the system ${\cal SA}$ in the initial state 
$\left\vert \Psi _{init}\right\rangle =\left\vert \psi\right\rangle \otimes \left\vert \varepsilon _{b}\right\rangle$, where 
$\left\vert \psi\right\rangle$ is an arbitrary state of ${\cal S}$ and 
$\left\vert \varepsilon _{b}\right\rangle$ is the (non-degenerate) ground state of $H^{\left( b\right) }$. Then $\left\vert \Psi _{init}\right\rangle$ 
is the ground state of the initial Hamiltonian $\mathbbm{1}\otimes H^{\left( b\right) }$. By applying the adiabatic theorem~\cite{Messiah:book,Sarandy:04} , 
a sufficiently slowing-varying evolution of $H(t)$ will drive the system (up to a phase) to the final state 
\begin{equation}
\left\vert \Psi_{final}\right\rangle =\sum\nolimits_{k}P_{k}\left\vert \psi \right\rangle \otimes \left\vert \varepsilon _{k}\right\rangle,
\label{psi-final}
 \end{equation}
where $\left\vert \varepsilon_{k}\right\rangle$ is the ground state of $H_{k}^{\left( f\right) }$~\cite{Hen:15} . 

\subsubsection{Single-qubit unitaries and controlled two-qubit gates}
We can perform a single-qubit unitary transformation through a general rotation of an angle $\phi$ around a direction $\hat{n}$ on the Bloch sphere.
In this direction, we begin by preparing the system ${\cal SA}$, taken here as two qubits, in the initial state $\left\vert\psi \right\rangle \otimes \left\vert 0\right\rangle$, 
where $\{|0\rangle, |1\rangle\}$ are the computational states of the auxiliary system ${\cal A}$.  
Then, we let the system adiabatically evolve driven by the Hamiltonian~\cite{Hen:15} 
\begin{equation}
H\left( s\right) =\left\vert \hat{n}_{+}\right\rangle \left\langle \hat{n}%
_{+}\right\vert \otimes H_{0}\left( s\right) +\left\vert \hat{n}%
_{-}\right\rangle \left\langle \hat{n}_{-}\right\vert \otimes H_{\phi
}\left( s\right)  ,  \label{cqa.2.2}
\end{equation}%
where $H_{0}(s)$ and $H_{\phi}(s)$ are adiabatically-evolved Hamiltonians, whose effect will be restricted to the respective subspaces of the projectors 
$\left\vert \hat{n}_{\pm }\right\rangle \left\langle \hat{n}_{\pm}\right\vert  = \frac{1}{2}\left( 1\pm \hat{n}\cdot \vec{\sigma}\right)$, where $\hat{n}$ is a unitary vector on the 
Bloch sphere associated with ${\cal S}$ and $\vec{\sigma}=\left( \sigma _{x},\sigma _{y},\sigma _{z}\right)$, with $\{\sigma_i\}$ denoting the set of Pauli matrices. The Hamiltonians 
are taken as  $H_{\xi }\left( s\right) = -\omega \hbar \left\{\sigma _{z}\cos (\theta_0 s) + \sin(\theta_0 s)  \left[ \sigma _{x}\cos \xi+\sigma _{y}\sin \xi \right] \right\}$, where ${\xi} \in \{ 0, \phi \}$,  
$\omega\hbar$ sets the energy scale ($\omega>0$), $\theta _{0}$ is a constant parameter, and $s=t/\tau$ denotes a dimensionless (parametrized) time, with $\tau$ the total time of evolution. Note then that 
\begin{eqnarray}
H_{0}\left( s\right) &=& 
-\omega \hbar \left\{\sigma _{z}\cos \left(\theta_0 s\right) +\sigma _{x}\sin \left(\theta_0 
s\right)\right\}, \nonumber \\ 
H_{\phi }\left( s\right) &=&  -\omega \hbar \left\{\sigma _{z}\cos (\theta_0 s) + \sin(\theta_0 s)  \left[ \sigma _{x}\cos \phi+\sigma _{y}\sin \phi \right] \right\}. \nonumber \\
\label{pw-H}
\end{eqnarray} 
By writing the initial state of ${\cal SA}$ as $\left\vert \Psi _{init}\right\rangle =\left\vert \psi
\right\rangle \otimes |0\rangle = \left(\alpha \left\vert \hat{n}_{+}\right\rangle +\beta \left\vert \hat{n}_{-}\right\rangle\right)\otimes |0\rangle$, where $|\psi\rangle$ is an arbitrary (not necessarily known)   
qubit state, the final state $|\Psi_{final}\rangle$ follows from 
Eq.~(\ref{psi-final}), i.e.  
$\left\vert \Psi _{final}\right\rangle =\alpha \left\vert \hat{n}%
_{+}\right\rangle \otimes\left\vert \varepsilon^{-}_{0}\right\rangle +\beta \left\vert \hat{n}%
_{-}\right\rangle \otimes\left\vert \varepsilon^{-}_{\phi }\right\rangle$. Note that $\left\vert \varepsilon_{\xi }^{-}\right\rangle $ is the ground state of $H_{\xi }\left( s\right)$, reading  
$\left\vert \varepsilon_{\xi }^{-}\right\rangle =\cos \left( \theta _{0}s/2\right) \left\vert 0\right\rangle
+e^{i\xi }\sin \left( \theta _{0}s/2\right) \left\vert 1\right\rangle$, with ${\xi} \in \{ 0, \phi \}$. An equivalent form of writing $\left\vert \Psi _{final}\right\rangle$ is given by 
\begin{eqnarray}
\left\vert \Psi _{final}\right\rangle &=&\cos \left( \theta _{0}/2\right) \left(
\alpha \left\vert \hat{n}_{+}\right\rangle +\beta \left\vert \hat{n}%
_{-}\right\rangle \right) \otimes \left\vert 0\right\rangle \nonumber \\ 
&&+\sin \left( \theta
_{0}/2\right) \left( \alpha \left\vert \hat{n}_{+}\right\rangle +e^{i\phi
}\beta \left\vert \hat{n}_{-}\right\rangle \right) \otimes \left\vert 1\right\rangle . \label{Itay.1.2}
\end{eqnarray}
Hence, we have a probabilistic implementation of the rotated target state $\alpha \left\vert \hat{n}_{+}\right\rangle +e^{i\phi }\beta \left\vert \hat{n}_{-}\right\rangle$ for an arbitrary angle $\phi$ around an arbitrarily 
chosen axis $\hat{n}$, with probability $\sin ^{2}\left( \theta _{0}/2\right)$. In particular, this probability approximates to one by taking $\theta _{0}\approx \pi$. 

In order to perform controlled rotations of a qubit by an angle $\phi$ around a direction $\hat{n}$, the starting point will be to take the subsystem ${\cal S}$ as a two-qubit system and keeping ${\cal A}$ as a single 
auxiliary qubit. The Hamiltonian is now chosen to be
\begin{eqnarray}
H\left( s\right)  &=&\left( \left\vert 0,\hat{n}_{+}\right\rangle
\left\langle 0,\hat{n}_{+}\right\vert +\left\vert 0,\hat{n}_{-}\right\rangle
\left\langle 0,\hat{n}_{-}\right\vert +\left\vert 1,\hat{n}_{+}\right\rangle
\left\langle 1,\hat{n}_{+}\right\vert \right) \otimes H_{0}\left( s\right)  \nonumber \\
&&+\left\vert 1,\hat{n}_{-}\right\rangle \left\langle 1,\hat{n}%
_{-}\right\vert \otimes H_{\phi }\left( s\right) , \label{itayDuplo}
\end{eqnarray}%
which will govern the evolution of the initial composite state 
$\left\vert \Psi _{init}\right\rangle = \left\vert \psi \right\rangle \otimes |0\rangle$, with 
$ \left\vert \psi \right\rangle = \alpha \left\vert 0,\hat{n}_{+}\right\rangle +\beta\left\vert 0,\hat{n}_{-}\right\rangle +\gamma \left\vert 1,\hat{n}_{+}\right\rangle +\delta \left\vert 1,\hat{n}_{-}\right\rangle$. 
From Eq.~(\ref{psi-final}), the final state of the subsystem ${\cal S}$ in the limit $\theta _{0}\rightarrow \pi $ is now the controlled rotated vector $\left\vert \psi^{Rot}\right\rangle =\alpha \left\vert 0,\hat{n}%
_{+}\right\rangle +\beta \left\vert 0,\hat{n}_{-}\right\rangle +\gamma\left\vert 1,\hat{n}_{+}\right\rangle +e^{i\phi }\delta \left\vert 1,\hat{n}_{-}\right\rangle$. By combining controlled rotations with the 
single-qubit unitaries discussed above, it is possible to design universal sets of quantum gates through an adiabatic implementation.  

\section{Results}
In this Section we present the main results of this work. We start  by generalizing the adiabatic implementation of quantum gates proposed in 
Ref.~\cite{Hen:15} for n-qubit controlled gates. 
Even though n-qubit controlled gates can be decomposable into one and two-qubit gates (see, e.g. Refs.~\cite{Barenco:95,Liu:08}), this 
implementation implies 
in an extended class of adiabatic universal gates, e.g.  the 
set $\{$Toffoli $,$ Hadamard$\}$~\cite{Shi:03,Aharonov:03} . Then, we will derive the main result of this work, which is 
a shortcut for general adiabatic circuits through constant 
counter-diabatic Hamiltonians, which implies in the possibility of fast analog implementations of quantum circuits.  Moreover, we will present an analysis 
of the quantum speed limit in the context of the energetic cost of the superadiabatic circuit.  

\subsection{Adiabatic $n$-controlled gates}
In order to implement $n$-controlled gates, we define the subsystem ${\cal S}$ as an $(n+1)$-qubit system, with the first $n$ qubits used as the control register and the last qubit taken as the target register. For the auxiliary 
system ${\cal A}$, we keep it as a single qubit.  Then, we take the initial state as $\left\vert \Psi _{init}\right\rangle =\left\vert \psi\right\rangle \otimes \left\vert 0\right\rangle$, with the subsystem ${\cal S}$ 
described by
\begin{equation}
\left\vert \psi \right\rangle =\sum\nolimits_{k_{1},\ldots
,k_{n},\epsilon }\gamma _{k_{1},\ldots ,k_{n},\epsilon }\left\vert
k_{1},\ldots ,k_{n},\hat{n}_{\epsilon }\right\rangle  , \label{AdG.1.4}
\end{equation}%
where $\gamma _{k_{1},\ldots ,k_{n},\epsilon }$ are complex amplitudes, $k_{l} \in \left\{ 0,1\right\}$, $\epsilon =\left\{ \pm \right\}$, and $\hat{n}$ is an arbitrary axis in the Bloch sphere. 
Here we have written the target qubit in the basis $\left\{ \left\vert \hat{n}_{\pm }\right\rangle \right\}$, leaving the remaining qubits of ${\cal S}$ in the computational basis. 
For simplicity, we will write the states in its decimal
representation, i.e. 
\[
\sum\nolimits_{k_{1},\ldots ,k_{n},\epsilon }\gamma _{k_{1},\ldots
,k_{n},\epsilon }\left\vert k_{1},\ldots ,k_n,\hat{n}_{\epsilon }\right\rangle
\rightarrow \sum_{m=0}^{N-1}\sum_{\epsilon=\pm }\gamma
_{m,\epsilon }\left\vert m,\hat{n}_{\epsilon }\right\rangle ,
\]%
where $N=2^{n}$. Then, we let the system evolve driven by the Hamiltonian
\begin{eqnarray}
H\left( s\right)  &=&\left[ 
\sum_{n=0}^{N-2}\sum_{\epsilon=\pm }\left\vert n,\hat{n}%
_{\epsilon }\right\rangle \left\langle n,\hat{n}_{\epsilon }\right\vert %
+ \left\vert N-1,\hat{n}_{+}\right\rangle
\left\langle N-1,\hat{n}_{+}\right\vert \right] \otimes H_{0}\left( s\right)   \nonumber \\
&&+\left\vert N-1,\hat{n}_{-}\right\rangle \left\langle N-1,\hat{n}%
_{-}\right\vert \otimes H_{\phi }\left( s\right) .  \label{AdG.1.6}
\end{eqnarray}%
We note that the rotation of the target qubit is expected to be applied if the
state of the control system is $\left\vert N-1\right\rangle$. Then, if the Hamiltonian is sufficiently 
slowly-varying so that we can apply the adiabatic theorem, 
the system will achieve the final state%
\begin{eqnarray}
\left\vert \Psi _{final}\right\rangle 
&=&\sum_{m=0}^{N-2}\sum_{\epsilon=\pm }\gamma _{m,\epsilon
}\left\vert m,\hat{n}_{\epsilon }\right\rangle \otimes \left\vert \varepsilon^{-}_{0}\right\rangle
+\gamma _{N-1,+}\left\vert N-1,\hat{n}_{+}\right\rangle \otimes \left\vert
\varepsilon^{-}_{0}\right\rangle   \nonumber \\
&&+\gamma _{N-1,-}\left\vert N-1,\hat{n}_{-}\right\rangle \otimes \left\vert \varepsilon^{-}_{\phi
}\right\rangle ,  \label{AdG.1.8}
\end{eqnarray}%
where $\left\vert \varepsilon^{-}_{\xi }\right\rangle $ is defined as $\left\vert \varepsilon^{-}_{\xi
}\right\rangle =\cos \left( \theta _{0}/2\right) \left\vert 0\right\rangle
+e^{i\xi }\sin \left( \theta _{0}/2\right) \left\vert 1\right\rangle$ ($\xi \in \{0,\phi\}$). 
An equivalent form of writing Eq.~(\ref{AdG.1.8}) is
\begin{equation}
\left\vert \Psi _{final}\right\rangle =\cos \left( \theta _{0}/2\right)
\left\vert \psi\right\rangle \otimes \left\vert 0\right\rangle +\sin \left(
\theta _{0}/2\right) \left\vert \psi^{Rot}\right\rangle \otimes \left\vert
1\right\rangle ,  \label{AdG.1.9}
\end{equation}%
with
\begin{eqnarray}
\left\vert \psi^{Rot}\right\rangle 
&=&\sum\nolimits_{m=0}^{N-2}\sum_{\epsilon=\pm }\gamma _{m,\epsilon
}\left\vert m,\hat{n}_{\epsilon }\right\rangle \nonumber \\
&&+ \left\vert N-1\right\rangle \otimes \left( \gamma _{N-1,+}\left\vert \hat{n}_{+}\right\rangle   
+\gamma _{N-1,-}e^{i\phi }\left\vert \hat{n}_{-}\right\rangle \right) . 
\label{AdG.1.10}
\end{eqnarray}
Thus, by performing a measurement over the auxiliary qubit, we find the rotated state $\left\vert \psi^{Rot}\right\rangle$ with probability 
$\sin ^{2}\left( \theta_{0}/2\right)$. As in the previous case of a rotation controlled by one qubit, this probability can be enhanced to one in the 
limit $\theta _{0}\rightarrow \pi$. Indeed, this state implies in a rotation of the target qubit in ${\cal S}$ conditioned by the state 
$\left\vert N-1\right\rangle \equiv |1 \cdots 1\rangle$ of the control register in ${\cal S}$. 
An application of this scheme is the adiabatic implementation of the Toffoli gate, which constitutes an unitary operation implementing an 
$X$ gate over the target qubit if all control qubits are in the state $1$, with no effect if any qubit of the control register is in
the state $0$. This can be easily achieved here by a rotation of an angle $\pi$ around of the direction x, therefore choosing 
$\phi =\pi $ and $\left\vert \hat{n}_{\pm }\right\rangle =\left\vert \pm \right\rangle$, with $\left\vert \pm \right\rangle$ denoting the eigenstates of $\sigma_x$.

\subsection{Shortcut to adiabaticity via counter-diabatic driving}
Let us introduce now a shortcut to general CAE through the superadiabatic approach. This will allow for fast piecewise implementation of 
quantum gates, whose evolution time will not be constrained by the adiabatic theorem. We begin by defining the 
evolution operator~\cite{Berry:09}%
\begin{equation}
U\left( t\right) =\sum_{n}e^{-\frac{i}{\hbar }\int_{0}^{t}dt^\prime
E_{n}\left( t^\prime \right) }e^{-\int_{0}^{t}dt^\prime \left\langle n|\partial
_{t^\prime }n\right\rangle }\left\vert n(t)\right\rangle \left\langle
n(0)\right\vert  , \label{s.1.1}
\end{equation}
which leads an initial state $|\Psi(0)\rangle = |n(0)\rangle$ into an evolved state $\left\vert \Psi \left( t\right) \right\rangle$ given by 
\begin{equation}
\left\vert \Psi \left( t\right) \right\rangle =e^{-\frac{i}{%
\hbar }\int_0^{t}d t^\prime E_{n}\left( t^\prime \right) }e^{-%
\int_0^{t}dt^\prime \left\langle n|\partial _{t^\prime }n\right\rangle }\left\vert n(t)\right\rangle ,\label{xx}
\end{equation}
where $\left\vert n\right\rangle =\left\vert n\left( t\right) \right\rangle $
are the eigenvectors of the adiabatic Hamiltonian. Note that this evolution mimics the adiabatic behavior. The Hamiltonian that guides the evolution of the system is the 
{\it superadiabatic} Hamiltonian, which reads
\begin{equation}
H_{SA}\left( t\right) =H\left( t\right) +H_{CD}\left( t\right) ,  \label{sfa.1.1}
\end{equation}%
where the additional term $H_{CD}\left( t\right) $ is the \textit{%
counter-diabatic} Hamiltonian
\begin{equation}
H_{CD}\left( t\right) =i\hbar \sum_{n}\left( \left\vert \partial
_{t}n\right\rangle \left\langle n\right\vert +\left\langle \partial
_{t}n|n\right\rangle \left\vert n\right\rangle \left\langle n\right\vert
\right) .  \label{sfa.1.2}
\end{equation}%
Therefore, a superadiabatic implementation of a dynamical evolution involves  
the knowledge of the eigenstates of the adiabatic Hamiltonian $H\left( t\right)$, which limits the direct application of the 
superadiabatic approach in quantum computation. For instance, by adopting the original AQC approach~\cite{Farhi:01} , 
superadiabatic implementations seem prohibitive, since the whole set of eigenlevels of a many-body Hamiltonian is required. In a similar context, 
counter-diabatic driving protocols based on realizable settings have been investigated for assisted evolutions in quantum critical phenomena~\cite{Campo:12,Campo:13,Saberi:14} . 
Here, as we shall see, the superadiabatic implementation of universal quantum circuits in the hybrid approach can be promptly achieved, 
since we deal with the eigenspectrum of piecewise Hamiltonians, which act over a few qubits.  It is then appealing to 
formulate a superadiabatic theory to CAE and then to specify it to the implementation of universal sets of quantum gates. 
Let us begin by establishing the complete set of eigenstates of the Hamiltonian $H\left( t\right)$ provided by Eq.~(\ref{sce.1.2}). 
To this end, consider the eigenvalue equation to each Hamiltonian $H_{k}\left( t\right)$ given by%
\begin{equation}
H_{k}\left( t\right) \left\vert \varepsilon_{k}^{i}\left( t\right)
\right\rangle =E_{k}^{i}\left( t\right) \left\vert \varepsilon_{k}^{i}\left( t\right) \right\rangle  , \label{sce1.3}
\end{equation}%
with $i=1,\cdots, d_{\cal A}$. By defining the projectors $P_{k}$ in Eq.~(\ref{sce.1.2}) as $P_{k}=\left\vert \lambda _{k}\right\rangle \left\langle \lambda
_{k}\right\vert$, with $\left\langle \lambda _{k}|\lambda _{k^{\prime
}}\right\rangle =\delta _{kk^{\prime }}$ and $k=1,\cdots, d_{\cal S}$, we can write the complete set of eigenstates of $H\left( t\right)$ as
\begin{equation}
\left\vert \gamma^{i}_{l}\left( t\right) \right\rangle =\left\vert \lambda_{l}\right\rangle \otimes \left\vert \varepsilon_{l}^{i}\left( t\right) \right\rangle ,
\label{sce1.5}
\end{equation}%
 such that
$H\left( t\right) \left\vert \gamma^{i}_{l}\left( t\right) \right\rangle
=E_{l}^{i}\left( t\right) \left\vert \gamma^{i}_{l}\left( t\right)
\right\rangle$. Indeed, from Eq.~(\ref{sce.1.2}), we have $H\left( t\right) \left\vert \gamma^{i}_{l}\left( t\right) \right\rangle = \left[\sum\nolimits_{k}P_{k}\otimes H_{k}\left( t\right)\right] 
\left[\left\vert \lambda_{l}\right\rangle \otimes \left\vert \varepsilon_{l}^{i}\left( t\right) \right\rangle\right] =%
E_{l}^{i}\left( t\right) \left\vert \gamma^{i}_{l}\left( t\right)\right\rangle$. 
Note that each projector $P_k$ is associated with a Hamiltonian $H_k$. For instance, for the adiabatic implementation of $n$-controlled gates, we have defined the 
Hamiltonian $H$ in Eq.~(\ref{AdG.1.6}) by linking the set    
$\{|m,n_{\pm}\rangle \langle m,n_{\pm}| \,(m=0,\cdots, N-2) , \,\, |N-1, n_{+}\rangle \langle N-1,n_{+} |\}$ with $H_0$ and by linking the remaining projector 
$|N-1, n_{-}\rangle \langle N-1,n_{-} |$ with $H_\phi$. The next step is 
to obtain the counter-diabatic Hamiltonian $H_{CD}\left( t\right)$ that implements the shortcut to the adiabatic evolution of 
$H\left( t\right)$. In this direction, we use the eigenstates of $H\left( t\right)$ as given by Eq.~(\ref{sce1.5}). Then, we get
\begin{equation}
H_{CD}\left( t\right) =i\hbar \sum\nolimits_{i}\sum\nolimits_{l}\left[
\left\vert \partial _{t}\gamma^{i}_{l}\right\rangle \left\langle \gamma^{i}
_{l}\right\vert +\left\langle \partial _{t}\gamma^{i}_{l}\,\,\left\vert \frac{}{}\hspace{-0.1cm}\right. \gamma^{i}
_{l}\right\rangle \left\vert \gamma^{i} _{l}\right\rangle \left\langle \gamma^{i}
_{l}\right\vert \right]  ,  \label{sce1.5.a}
\end{equation}%
with $\left\vert \gamma^{i} _{l}\right\rangle \equiv \left\vert \gamma^{i} _{l}\left(
t\right) \right\rangle $. Therefore  
\begin{eqnarray}
H_{CD}\left( t\right) &=&i\hbar \sum_{l } \left[ |\lambda_l\rangle \langle \lambda_l | \otimes  \sum_{i}\left(
\left\vert \partial _{t}\varepsilon^{i}_{l}\right\rangle \left\langle \varepsilon^{i}_{l}\right\vert +\left\langle \partial _{t}\varepsilon^{i}_l \,\,\left\vert \frac{}{}\hspace{-0.1cm} \right.  
\varepsilon^{i}_{l}\right\rangle \left\vert \varepsilon^{i}_{l}\right\rangle \left\langle \varepsilon^{i}_{l}\right\vert \right) \right] \nonumber \\
&=& \sum\nolimits_{l}\left[ P_{l}\otimes H_{l}^{CD}\left(
t\right) \right] ,  \label{sce1.5.b}
\end{eqnarray}%
where $\left\vert \varepsilon^{i} _{l}\right\rangle \equiv \left\vert \varepsilon^{i} _{l}\left(
t\right) \right\rangle $ and $H_{l}^{CD}$ is the counter-diabatic Hamiltonian to be associated with the piecewise adiabatic contribution $H_{l}\left( t\right)$ acting over subsystem ${\cal A}$, 
which reads
\begin{equation}
H_{l}^{CD}\left( t\right) =i\hbar \sum\nolimits_{i}\left[ \left\vert
\partial _{t}\varepsilon^{i} _{l}\right\rangle \left\langle \varepsilon^{i} _{l}\right\vert +\left\langle \partial _{t}\varepsilon^{i} _{l}\,\,\left\vert \frac{}{}\hspace{-0.1cm}\right. 
\varepsilon^{i} _{l}\right\rangle \left\vert \varepsilon^{i} _{l}\right\rangle
\left\langle \varepsilon^{i} _{l}\right\vert \right]  . \label{sce1.6}
\end{equation}%
Hence, from Eq.~(\ref{sfa.1.1}), we can implement the shortcut dynamics through the superadiabatic Hamiltonian
\begin{equation}
H_{SA}\left( t\right) =\sum\nolimits_{k}P_{k}\otimes H_{k}^{SA}\left(
t\right) ,  \label{sce1.7}
\end{equation}%
where $H_{k}^{SA}\left( t\right)\equiv H_k\left( t\right) +H_{k}^{CD}\left( t\right)  $ is the piecewise superadiabatic Hamiltonian. Note that the cost of
performing superadiabatic evolutions requires the knowledge of the eigenvalues and eigenstates 
of $H_{l}\left( t\right)$. For the implementation of general $n$-controlled gates, this is a Hamiltonian acting over a single qubit, which is independent of 
the circuit complexity. Moreover, we can show that, for an arbitrary 
$n$-controlled quantum gate, the counter-diabatic Hamiltonians $H_{\xi}^{CD}$ (${\xi} \in \{ 0, \phi \}$) associated with shortcuts to 
adiabatic evolutions driven by $H_{\xi }\left( s\right) = -\omega \hbar \left\{\sigma _{z}\cos (\theta_0 s) + \sin(\theta_0 s)  \left[ \sigma _{x}\cos \xi+\sigma _{y}\sin \xi \right] \right\}$, 
with $s = t / \tau$, are {\it time-independent} operators given by
 \begin{equation}
H_{\xi }^{CD}=\hbar \frac{\theta _{0}}{2\tau}\left[ \sigma _{y}\cos \xi - \sigma
_{x}\sin \xi \right] .  \label{sce1.7.ad}
\end{equation}%
Eq.~(\ref{sce1.7.ad}) shows that the implementation of the shortcut can be achieved with a very simple assistant Hamiltonian in the quantum dynamics. Its proof is provided in 
Section {\it Methods}.  

\subsection{Quantum speed limit}
It is expected that the shortcut via a counter-diabatic Hamiltonian is faster than
the evolution via an adiabatic Hamiltonian, but how much faster can it be? To answer this
question, we shall take a lower bound to the time evolution in quantum 
dynamics as provided by the {\it quantum speed limit} (QSL). 
We will consider a closed quantum system evolving between arbitrary pure states $\left\vert
\Psi \left({0}\right)\right\rangle $ and $\left\vert \Psi \left( \tau \right)
\right\rangle $. Since the evolution is 
driven by a time-dependent superadiabatic Hamiltonian $H_{SA}(t)$, we will take the generalized Margolus-Levitin bound~\cite{Margolus:98}   
derived by Deffner and Lutz~\cite{Deffner:13} , which reads
\begin{equation}
\tau \geq \hbar \frac{\left\vert \cos \mathcal{L}\left( \Psi \left( 0\right)  ,\Psi \left( \tau\right)  \right) -1\right\vert }{E_{\tau }} ,
\label{qsl.1}
\end{equation}%
where $\mathcal{L}\left( \Psi \left( 0\right) ,\Psi \left( \tau\right) \right) =\arccos \left[
\left\vert \left\langle \Psi _{0}|\Psi \left( \tau \right) \right\rangle
\right\vert \right] $ is the Bures metric for pure states \cite{Nielsen:book} and%
\begin{equation}
E_{\tau }=\frac{1}{\tau }\int_{0}^{\tau }dt\left\vert \left\langle \Psi(0)|H_{SA}\left( t\right) |\Psi \left( t\right) \right\rangle \right\vert .
\label{qsl.2}
\end{equation}
For superadiabatic evolutions, the initial state $\left\vert \Psi \left(
0\right) \right\rangle = |\gamma_0(0)\rangle$ evolves to $\left\vert \Psi \left(
t\right) \right\rangle = |\gamma_0(t)\rangle$, where $|\gamma_0(t)\rangle$ denotes the instantaneous ground state of the adiabatic Hamiltonian $H(t)$. 
By using the parametrized time $s=t/\tau$, we can show from Eqs.~(\ref{qsl.1}) and~(\ref{qsl.2}) that the total time $\tau$ that mimics 
the adiabatic evolution within the superadiabatic approach can be reduced to an arbitrary small value. More specifically, the addition of a 
counter-diabatic Hamiltonian implies into the QSL bound 
\begin{equation}
\eta \, \tau \omega +  \chi \, \ge \, \hbar \left\vert \cos \mathcal{L}\left( \Psi \left( 0\right)  ,\Psi \left( \tau\right)  \right) -1\right\vert  ,
\label{qsl.final}
\end{equation}
with $\eta  > 0$ and $\chi\ge \hbar \left\vert \cos \mathcal{L}\left( \Psi \left( 0\right)  ,\Psi \left( \tau\right)  \right) -1\right\vert$, as shown in Section {\it Methods}. 
Therefore, the QSL bound reduces to
\begin{equation}
 \tau\omega  \ge 0 \hspace{0.5cm} ( \,\, \forall \,\,\,  \left\vert \Psi \left(
0\right) \right\rangle \,\,\, , \,\,\, \left\vert \Psi \left(
\tau\right) \right\rangle \,\, ) , 
\label{qsl.final2}
\end{equation}
with $\tau$ and $\omega$ defined by the superadiabatic Hamiltonian $H_{SA}(t)$. 
This means that the superadiabatic implementation is compatible with an arbitrary reduction of the total time $\tau$, which 
holds {\it independently} of the boundary states $\left\vert \Psi \left( 0\right) \right\rangle$ and $\left\vert \Psi \left(
\tau\right) \right\rangle$. Naturally, a higher energetic cost is expected to be involved for a smaller evolution time $\tau$.
In particular, saturation of Eq.~(\ref{qsl.final2}) is achieved for either $\tau \rightarrow 0$ or $\omega \rightarrow 0$, with both cases implying in $\tau\omega \rightarrow 0$. 
Note that this limit is forbidden in the adiabatic regime for finite $\omega$, since the energy gap is proportional to $\hbar\omega$, which implies in an adiabatic time of the order 
$\tau_{ad} \propto 1/ \omega^n$, with $n \in \mathbb{N}^+$~~~\cite{Messiah:book,Teufel:03,Jansen:07,Sarandy:04} . Hence, Eq.~(\ref{qsl.final2}) leads to a flexible running time in a superadiabatic implementation, only limited by the energy-time complementarity.  
 
\subsection{The Energetic Cost}
Let us show now that time and energy are complementary resources in superadiabatic implementations of quantum evolutions. 
We shall define the energetic cost associated with a superadiabatic Hamiltonian through
\begin{equation}
\Sigma \left( \tau \right) =\frac{1}{\tau }\int_{0}^{\tau }\left\Vert 
{H}_{SA}\left( t\right) \right\Vert dt  , \label{cost1.1}
\end{equation}%
with ${H}_{SA}\left( t\right) $ given by %
Eq.~(\ref{sce1.7}) and the norm provided by the Hilbert-Schmidt norm $\left\Vert A\right\Vert =\sqrt{Tr\left[ A^{\dag }A%
\right] }$. Since ${H}_{SA}\left( t\right) $ is Hermitian, we can write%
\begin{eqnarray}
\Sigma \left( \tau \right) &=&\frac{1}{\tau }\int_{0}^{\tau }\sqrt{\text{Tr}%
\left[ {H}_{SA}^{2}\left( t\right) \right] }dt  \nonumber \\
&=&\frac{1}{\tau }\int_{0}^{\tau }\sqrt{\text{Tr}%
\left[ {H}^{2}\left( t\right) + {H}_{CD}^{2}\left( t\right) \right] }dt . \label{cost1.2}
\end{eqnarray}
To derive Eq.~(\ref{cost1.2}), we have used that ${\textrm{Tr}} \left( \left\{ H(t), \frac{}{} \hspace{-0.05cm}H_{CD}(t) \right\} \right) = 0$. This can be obtained by computing the trace in the eigenbasis of $H(t)$ and noticing that 
the expectation value of  $H_{CD}(t)$ taken in an eigenstate of $H(t)$ vanishes, i.e. 
$\left\langle \gamma_l^i (t) \right\vert H_{CD}(t) \left\vert \gamma_l^i(t)\right\rangle = 0$.
In particular, let us define the energetic cost to the adiabatic Hamiltonian as
\begin{equation*}
\Sigma _{0}\left( \tau \right) =\frac{1}{\tau }\int_{0}^{\tau }\sqrt{Tr\left[
{H}^{2}\left( t\right) \right] }dt . 
\end{equation*}%
Then, it follows that the energetic cost $\Sigma \left( \tau \right)$ in superadiabatic evolutions supersedes the energetic cost $\Sigma _{0}\left( \tau \right)$ for 
a corresponding adiabatic physical process. 
In order to evaluate 
$\Sigma \left( \tau \right)$ we adopt the basis of eigenstates of the adiabatic Hamiltonian ${H}\left( t\right) $. By using Eq.~(\ref{sce1.5}), this yields
\begin{equation}
\Sigma \left( \tau \right) 
=\frac{1}{\tau }\int_{0}^{\tau }\sqrt{%
\sum\nolimits_{l=1}^{d_{{\cal S}}}  \sum\nolimits_{m=1}^{d_{{\cal A}}}\left[ E_{l}^{m}\left( t\right) ^{2}+\hbar
^{2}\mu _{l}^{m}\left( t\right) \right]   }dt
,  \label{cost1.4}
\end{equation}
where $E_{l}^{m}\left( t\right) $ are the energies of the adiabatic Hamiltonian $%
H_{l}\left( t\right) $ and 
\begin{equation}
\mu _{l}^{m}\left( t\right) = \left\langle \partial_t{\gamma} _{l}^{m}\left( t\right) |\partial_t{\gamma}_{l}^{m}\left( t\right)
\right\rangle - \left\vert
\left\langle \gamma _{l}^{m}\left( t\right) |\partial_t{\gamma}_{l}^{m}\left( t\right)
\right\rangle \right\vert ^{2}.
\end{equation} 
In order to analyze the energetic cost as provided by Eq.~(\ref{cost1.4}) for superadiabatic qubit rotation gates, we set $d_{{\cal S}}=d_{{\cal A}}=2$ and $E_{l}^{m}\left( s\right)
=\left( -1\right) ^{m+1}\hbar \omega $ $(\forall l)$. Moreover, by using Eq.~(\ref{sce1.5}), we obtain 
$\mu _{l}^{m}\left( t\right) = \left\langle \partial_t{\varepsilon} _{l}^{m}\left( t\right) |\partial_t{\varepsilon}_{l}^{m}\left( t\right)
\right\rangle - \left\vert
\left\langle \varepsilon _{l}^{m}\left( t\right) |\partial_t{\varepsilon}_{l}^{m}\left( t\right)
\right\rangle \right\vert ^{2}$, which leads to 
$\mu _{l}^{m}\left( s\right) =\theta _{0} ^{2}
/4\tau^{2} $ [See Eqs.~(\ref{cqa.2.5a}) and (\ref{cqa.2.5b}) in Section {\it Methods}]. Hence
\begin{equation}
\Sigma\left( \tau \right)=2\sqrt{1+\frac{\theta _{0} ^{2}}{4\left( \tau \omega \right) ^{2}}}%
\hbar \omega .  \label{cost1.5}
\end{equation}

\begin{figure}[!ht]
\centering
\includegraphics[scale=0.4]{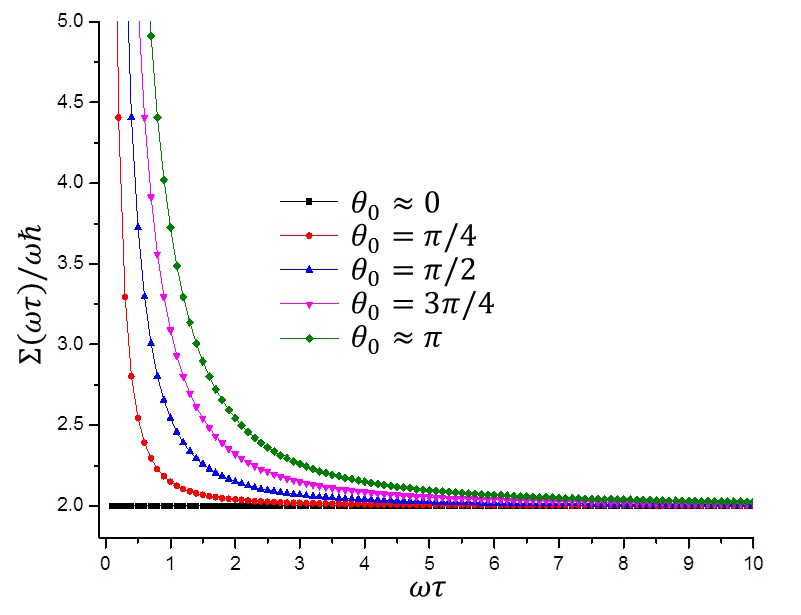}
\caption{Energetic cost in unities of $\hbar \omega$ as a function of $\omega \tau$ for different values of $\theta _{0}$. }
\label{graph1}
\end{figure}

We illustrate the behavior of $\Sigma\left( \tau \right)$ in Fig.~\ref{graph1}, where it is apparent that the energetic cost increases inversely proportional to the total time of evolution. 
In particular, note also that, for a fixed energetic cost, the optimal choice $\ \theta _{0}\rightarrow \pi $ requires a longer evolution. This is because of the fact that, in this case, 
the final state associated with the auxiliary qubit is orthogonal to its initial state, so it is farther in the Bloch sphere. In the more general case of controlled gates, the analysis is similar 
as in the case of single-qubit gates. However, we must take into account the number of projectors composing the set $\{P_k\}$. More specifically, the sum over $l$ in Eq.~(\ref{cost1.4}) 
shall run over $1$ to $4$, which is the number of projectors over the subsystem ${\cal S}$. Thus we can show that energetic cost $\Sigma^{CG}$ to implement controlled gates is 
$\Sigma^{CG}=\sqrt{2} \Sigma$.

\section{Methods}

\subsection{Time-independent counter-diabatic Hamiltonians for $n$-controlled gates}
Let us explicitly design here the superadiabatic implementation of controlled evolutions for piecewise Hamiltonians $H_{\xi}(s)$ as  
provided by Eqs.~(\ref{pw-H}). To this end, consider the eigenvalue equation%
\begin{equation}
H_{\xi }\left( s\right) \left\vert \varepsilon_{\xi }^{i}\left( s\right) \right\rangle
=E_{\xi }^{i}\left( s\right) \left\vert \varepsilon_{\xi }^{i}\left( s\right)
\right\rangle  , \label{cqa.2.4}
\end{equation}%
with $\xi =\left\{ 0,\phi \right\} $, where%
\begin{eqnarray}
\left\vert \varepsilon_{\xi }^{+}\left( s\right) \right\rangle  &=&-\sin 
\frac{\theta_0 s }{2}\left\vert 0\right\rangle +e^{i\xi }\cos \frac{\theta_0 s }{2}\left\vert
1\right\rangle  , \label{cqa.2.5a} \\
\left\vert \varepsilon_{\xi }^{-}\left( s\right) \right\rangle  &=&\cos 
\frac{\theta_0 s }{2}\left\vert 0\right\rangle +e^{i\xi }\sin \frac{\theta_0 s }{2}\left\vert
1\right\rangle  . \label{cqa.2.5b}
\end{eqnarray}
From Eq.~(\ref{sce1.5}), it follows that the 
eigenstates for the adiabatic Hamiltonian $H\left( s\right) $ governing the composite system ${\cal{SA}}$ 
are given by the sets $\left\{ \left\vert
\gamma^{i}_0\left( s\right) \right\rangle =\left\vert \hat{n}_{+}\right\rangle \otimes
\left\vert \varepsilon_{0}^{i}\left( s\right) \right\rangle \right\} $ and $\left\{
\left\vert \gamma^{i}_{\phi }\left( s\right) \right\rangle =\left\vert \hat{n}%
_{-}\right\rangle \otimes \left\vert \varepsilon_{\phi }^{i}\left( s\right) \right\rangle
\right\} $ associated with the set of eigenvalues $\left\{ E^{i}_0\left( s\right)
\right\} $ and $\left\{ E^{ i}_\phi\left( s\right) \right\} $, respectively.
By evaluating the eigenvalues of $H_{0}\left( s\right) $ and $%
H_{\phi }\left( s\right) $, we obtain that their spectra are equal, being provided 
by $E_{0}^{\pm }=E_{\phi }^{\pm }=\pm \omega \hbar $. Thus,  
$H\left( s\right) $ exhibits doubly degenerate levels, with $\left\{ \left\vert
\gamma^{+}_0\left( s\right) \right\rangle ,\left\vert \gamma^{ +}_\phi\left( s\right)
\right\rangle \right\} $ and $\left\{ \left\vert \gamma^{-}_0\left( s\right)
\right\rangle ,\left\vert \gamma^{ -}_\phi\left( s\right) \right\rangle \right\} $
 associated with levels $E^{+}=\omega \hbar $ and $%
E^{-}=-\omega \hbar $, respectively. By using now Eqs.~(\ref{cqa.2.5a}) and 
(\ref{cqa.2.5b}), we obtain 
$\left\langle \partial _{t}\varepsilon^{i}_l \,\,\left\vert \frac{}{}\hspace{-0.1cm} \right.  
\varepsilon^{i}_{l}\right\rangle=0$, for any $i =\left\{ \pm \right\} $ and $\xi =\left\{ 0,\phi \right\} $. 
Then, from Eq.~(\ref{sce1.5.b}), we obtain that the counter-diabatic 
hamiltonian is 
$H_{\xi }^{CD}\left( s\right) =i\hbar \sum\nolimits_{i =\left\{ \pm
\right\} }\left\vert \partial _{t}\varepsilon_{\xi }^{i }\left( s\right)
\right\rangle \left\langle \varepsilon_{\xi }^{i }\left( s\right) \right\vert$, which leads to the 
time-independent counter-diabatic Hamiltonian given by Eq.~(\ref{sce1.7.ad}).
The extension to the case of $n$-controlled gates can be achieved as follows. 
From Eq.~(\ref{sce1.5}), the eigenstates of $H\left( s\right) $ read%
\begin{eqnarray*}
\left\vert \gamma_{0 m}^{\epsilon k}\left( s\right) \right\rangle  &=&\left\vert m,%
\hat{n}_{\epsilon }\right\rangle \otimes \left\vert \varepsilon_{0}^{k}\left( s\right)
\right\rangle ,  \\
\left\vert \gamma_{0 \, (N-1)}^{+k}\left( s\right) \right\rangle  &=&\left\vert N-1,\hat{n}%
_{+}\right\rangle \otimes \left\vert \varepsilon_{0}^{k}\left( s\right) \right\rangle , \\
\left\vert \gamma_{\phi \, (N-1)}^{-k}\left( s\right) \right\rangle  &=&\left\vert N-1,%
\hat{n}_{-}\right\rangle \otimes \left\vert \varepsilon_{\phi }^{k}\left( s\right)
\right\rangle ,
\end{eqnarray*}
where $m=\left\{ 0,\cdots,N-2\right\} $, $\epsilon,k=\left\{ \pm \right\} $ and $\xi =\left\{
0,\phi \right\} $. 
By computing the eigenvalues of $H\left( s\right) $,  we obtain that 
the spectrum of $H\left( s\right) $ is $\left( 2N\right) $-degenerate, with 
$\left\{\left\vert \gamma_{0 m}^{\epsilon +}\left( s\right) \right\rangle ,\left\vert
\gamma_{0 \, (N-1)}^{++}\left( s\right) \right\rangle ,\left\vert \gamma_{\phi \, (N-1) }^{-+}\left(
s\right) \right\rangle \right\} $ and $\left\{ \left\vert \gamma_{0 m}^{\epsilon
-}\left( s\right) \right\rangle ,\left\vert \gamma_{0 \, (N-1)}^{+-}\left( s\right)
\right\rangle ,\left\vert \gamma_{\phi \, (N-1) }^{--}\left( s\right) \right\rangle
\right\} $ associated with the levels $E^{+}=\omega \hbar $ and $%
E^{-}=-\omega \hbar $, respectively. By using these results into Eq.~(\ref{sce1.5.b}), we obtain 
that the counter-diabatic piecewise Hamiltonian $H_{\xi }^{CD}\left(
s\right) $ is given by Eq.~(\ref{sce1.7.ad}). Hence, the implementation any 
$n$-controlled gate is achieved through a time-independent counter-diabatic Hamiltonian.

\subsection{Quantum speed limit for superadiabatic evolutions}
Let us apply here the QSL bound to superadiabatic evolutions. 
By using the fact than the $\left\vert \Psi \left( t\right) \right\rangle $
evolves in the ground state $\left\vert \gamma _{0}\left( t\right)
\right\rangle $ of $H\left( t\right) $ and that $H_{SA}\left( t\right) $
is given by Eq.~(\ref{sfa.1.1}), we have%
\begin{eqnarray*}
E_{\tau } &=&\frac{1}{\tau }\int_{0}^{\tau }dt\left\vert \left\langle
\gamma _{0}\left( 0\right) |H_{SA}\left( t\right) |\gamma
_{0}\left( t\right) \right\rangle \right\vert  , \\	
&=&\frac{1}{\tau }\int_{0}^{\tau }dt\left\vert E _{0}\left(
t\right) \left\langle \gamma _{0}\left( 0\right) |\gamma
_{0}\left( t\right) \right\rangle +\left\langle \gamma _{0}\left(
0\right) |H_{CD}\left( t\right) |\gamma _{0}\left( t\right)
\right\rangle \right\vert ,
\end{eqnarray*}
where $E_0(t)$ is the instantaneous ground state energy of $H(t)$. 
Now we use Eq.~(\ref{sfa.1.2}) and the inequality $\int_{0}^{\tau
}dt\left\vert f\left( t\right) +g\left( x\right) \right\vert \le
\int_{0}^{\tau }dt\left\vert f\left( t\right) \right\vert +\int_{0}^{\tau
}dt\left\vert g\left( t\right) \right\vert $, which yields
\begin{eqnarray}
E_{\tau } &\le& \frac{1}{\tau }\int_{0}^{\tau }dt\left\vert E
_{0}\left( t\right) \left\langle \gamma _{0}\left( 0\right)
|\gamma _{0}\left( t\right) \right\rangle \right\vert +\frac{\hbar }{%
\tau }\int_{0}^{\tau }dt\left\vert \left\langle \gamma _{0}\left(
0\right) |\partial_t{\gamma}_{0}\left( t\right) \right\rangle \right\vert \nonumber \\  
&&+\frac{\hbar }{\tau }\int_{0}^{\tau }dt\left\vert \left\langle \gamma
_{0}\left( t\right) |\partial_t{\gamma}_{0}\left( t\right) \right\rangle
\left\langle \gamma _{0}\left( 0\right) |\gamma _{0}\left(
t\right) \right\rangle \right\vert .  \label{dqsl.1}
\end{eqnarray}%
By using the parametrized time $s=t/\tau $, we obtain  
\begin{equation*}
E_{\tau }\le \eta _{1}(s)+\frac{\left( \eta _{2}(s)+\eta _{3}(s)\right) }{\tau } ,
\end{equation*}%
where the parameters $\eta_i(s)$ ($i=1,2,3$) are given by $\eta _{1}(s)=\int_{0}^{1}ds\left\vert E _{0}\left(
s\right) \left\langle \gamma _{0}\left( 0\right) |\gamma
_{0}\left( s\right) \right\rangle \right\vert ,$ $\eta
_{2}(s)=\hbar \int_{0}^{1}ds\left\vert \left\langle \gamma _{0}\left( 0\right)
|\partial_s\gamma _{0}\left( s\right) \right\rangle \right\vert $ and $%
\eta _{3}(s)=\hbar \int_{0}^{1}ds\left\vert \left\langle \gamma _{0}\left(
s\right) |\partial_s\gamma _{0}\left( s\right) \right\rangle
\left\langle \gamma _{0}\left( 0\right) |\gamma _{0}\left(
s\right) \right\rangle \right\vert $. Since the ground state energy for the adiabatic Hamiltonian $H(s)$ in the case of 
$n$-controlled gates is $E_0(s)=-\omega \hbar$ [see Eqs.~(\ref{pw-H}) and (\ref{AdG.1.6})], we write $\eta_1(s) =  \omega \eta(s)$, with 
 $\eta(s) = \hbar \int_{0}^{1}ds\left\vert \left\langle \gamma _{0}\left( 0\right) |\gamma
_{0}\left( s\right) \right\rangle \right\vert$. Moreover, we define $\chi(s) \equiv \eta_2(s) + \eta_3(s)$. Then
\begin{equation}
\eta(s) \, \omega \tau + \chi(s) \geq \hbar \left\vert \cos 
\mathcal{L}\left( \gamma _{0}\left( 0\right) ,\gamma _{0}\left(
1\right) \right) -1\right\vert  . \label{dqsl.2}
\end{equation}%
Let us now analise the term $\chi(s)$. First, note that $\left\langle
\gamma _{0}\left( 0\right) |\partial_s\gamma _{0}\left( s\right)
\right\rangle =d_{s}\left(\left\langle \gamma _{0}\left( 0\right)
|\gamma _{0}\left( s\right) \right\rangle\right)$.  Then, we use that $\left\vert d_{s}\left\langle
\gamma _{0}\left( 0\right) |\gamma _{0}\left( s\right)
\right\rangle \right\vert \geq \left\vert d_{s}\left\vert \left\langle
\gamma _{0}\left( 0\right) |\gamma _{0}\left( s\right)
\right\rangle \right\vert \right\vert $ (see proof in Ref.~\cite{Deffner:13}), which yields
\begin{eqnarray}
\chi(s) &\ge& \eta _{2}(s)\geq \hbar \int_{0}^{1}ds\left\vert d_{s}\left\vert \left\langle
\gamma _{0}\left( 0\right) |\gamma _{0}\left( s\right)
\right\rangle \right\vert \frac{}{}\right\vert \nonumber \\
&\ge& \hbar \left\vert \int_{0}^{1}ds\left( d_{s}\left\vert \left\langle
\gamma _{0}\left( 0\right) |\gamma _{0}\left( s\right)
\right\rangle \right\vert \right) \right\vert \nonumber \\
&=& \hbar \left\vert \left\vert \left\langle
\gamma _{0}\left( 0\right) |\gamma _{0}\left( 1\right)
\right\rangle \right\vert -1  \right\vert ,
\end{eqnarray}%
where we have used the inequality $\int_{0}^{\tau }dt\left\vert f\left( t\right)
\right\vert \geq \left\vert \int_{0}^{\tau }dtf\left( t\right) \right\vert $.
From the definition of the Bures metric, we have $\left\vert \left\langle
\gamma _{0}\left( 0\right) |\gamma _{0}\left( 1\right)
\right\rangle \right\vert =\cos 
\mathcal{L}\left( \gamma _{0}\left( 0\right) ,\gamma _{0}\left(
1\right) \right) $. Hence,  $ \chi(s) \geq \hbar \left\vert \cos \mathcal{L}\left(
\psi \left( 0\right) ,\psi\left( \tau\right) \right)
-1\right\vert $, which implies into Eq.~(\ref{qsl.final2}).

\section{Discussion and conclusion}

We have proposed a scheme for implementing universal sets of quantum gates within the superadiabatic approach. In particular, we have shown that this can be achieved by applying a 
{\it time-independent} counter-diabatic Hamiltonian in the auxiliary qubit to induce fast controlled evolutions. Remarkably, this Hamiltonian is universal, holding both for performing single-qubit 
and $n$-controlled qubit gates. Therefore, a shortcut to the adiabatic implementation of quantum gates can be achieved through a rather simple mechanism. In particular, different sets of 
universal quantum gates can be designed by using essentially the same counter-diabatic Hamiltonian. Moreover, we have shown that the flexibility of the evolution time in a superadiabatic 
dynamics can be directly traced back from the QSL bound. In this context, the running time is only constrained by the energetic cost of the superadiabatic implementation, within a time-energy 
complementarity relationship. Implications of the superadiabatic approach under decoherence and a fault-tolerance analysis of superadiabatic circuits are further challenges of immediate interest. 
In a quantum open-systems scenario, there is a compromise between the time required by adiabaticity and the decoherence time of the quantum device. Therefore, the superadiabatic implementation may provide a direction to obtain an optimal running time for the quantum algorithm while keeping an inherent protection against decoherence. In turn, a basis for such 
development may be provided by the generalization of the superadiabatic theory for the context of open systems~\cite{Jing:13,Vacanti:14,Jing:15,Shi:14} . 
Concerning error-protection, it may also be fruitful the comparison of our approach with non-adiabatic holonomic quantum computation, where non-adiabatic geometric phases are used to 
perform universal quantum gates (see, e.g. recent proposals in Refs.~\cite{Feng:13,Xu:14} ).
Moreover, the behavior of correlations 
such as entanglement may also be an additional relevant resource for superadiabaticity applied to quantum computation. These investigations as well as experimental proposals 
for superadiabatic circuits are left for future research.     

\section*{Acknowledgements}

We are grateful to Itay Hen and Adolfo del Campo for useful discussions. 
M. S. S. thanks Daniel Lidar for his hospitality at the University of Southern California.  
We acknowledge financial support from the Brazilian agencies CNPq, CAPES, and FAPERJ. 
This work has been performed as part of the Brazilian National Institute of Science and 
Technology for Quantum Information (INCT-IQ).

\end{document}